\documentclass{article}

\def\abstract#1{\vskip 7mm 
        \begin{center}{\large Abstract}\\[3mm]
                \begin{minipage}[c]{13cm}
                        \small #1
                \end{minipage}
        \end{center}
}
\def\title#1{\begin{center}{\Large\bf #1}\end{center}}
\def\author#1{\vskip 5mm \begin{center}{#1}\end{center}}
\def\address#1{\begin{center}{\it #1}\end{center}}

\newcommand{\eqb}{\begin{equation}}
\newcommand{\eqe}{\end{equation}}
\newcommand{\eqab}{\begin{eqnarray}}
\newcommand{\eqae}{\end{eqnarray}}
\newcommand{\eqabnon}{\begin{eqnarray*}}
\newcommand{\eqaenon}{\end{eqnarray*}}

\newcommand{\eqref}[1]{(\ref{#1})}
\newcommand{\defeq}{:=}

\newcommand{\cd}[1]{\dot{#1}}

\newcommand{\Tne}{T_{\rm ne}}
\newcommand{\ene}{\varepsilon_{\rm ne}}
\newcommand{\pne}{p_{\rm ne}}
\newcommand{\sne}{s_{\rm ne}}

\newcommand{\sv}{\widetilde{\Pi}}

\newcommand{\inthb}{\beta_{\rm hb}}
\newcommand{\inths}{\beta_{\rm hs}}

\newtheorem{thm}{Theorem}

\makeatletter
\@ifundefined{lesssim}{\def\lesssim{\mathrel{\mathpalette\vereq<}}}{}
\@ifundefined{gtrsim}{}{}
\def\vereq#1#2{\lower3pt\vbox{\baselineskip1.5pt \lineskip1.5pt
\ialign{$\m@th#1\hfill##\hfil$\crcr#2\crcr\sim\crcr}}}
\makeatother

\begin{document}

\title{%
Relativistic Dissipative Accretion Flow onto Black Hole
}
\author{%
 Hiromi Saida$^{(a)}$\,\footnote{Email address: saida@daido-it.ac.jp}, 
 Rohta Takahashi$^{(b)}$\,\footnote{Email address: rohta@riken.jp} and
 Hiroki Nagakura$^{(c)}$\,\footnote{Email address: hiroki@heap.phys.waseda.ac.jp}
}
\address{%
 $^{(a)}$ Department of Physics, Daido University, Nagoya, Japan\\
 $^{(b)}$ Cosmic Radiation Laboratory, the Institute of Physical and Chemical Research, Saitama, Japan\\
 $^{(c)}$ Department of Science and Engineering, Waseda University, Tokyo, Japan
}

\abstract{
Dissipations, e.g. heat flow and bulk and shear viscosities, cause the transport of energy, momentum and angular momentum, which is the essence of accretion of matters onto celestial objects. 
Dissipations are usually described by the Fourier and Navier-Stokes laws (\emph{classic laws} of dissipations). 
However the classic laws result in an infinitely fast propagation of dissipations. 
In relativistic formulation, the classic laws of dissipations violate the causality, and hence no relativistic theory of accretion flow onto celestial object is formulated. 
In this short report, we summarize the causal dissipative hydrodynamics, so-called \emph{Extended Irreversible Thermodynamics} (EIT), with a supplemental comment of which the original works of EIT are not aware, and then show two theorems about relativistic dissipative flows around a Schwarzschild black hole. 
By these theorems, a significant property of EIT in contrast with classic laws of dissipations is clarified, and a dissipative instability of an exact solution of relativistic perfect fluid flow is also obtained. 
}

\section{Extended Irreversible Thermodynamics (EIT)}

We begin with summarizing the basis of perfect fluid and classic laws of dissipations (e.g. Fourier and Navier-Stokes laws) in order to clarify the basis and need for EIT. 
The perfect fluid is a phenomenology assuming the \emph{local equilibrium} of fluid, which requires that each fluid element is in a thermal equilibrium state. 
By the local equilibrium assumption, any fluid element in perfect fluid evolves adiabatically and no entropy production arises in the fluid element. 
This contradicts the dissipative phenomena which are essentially the irreversible and entropy producing processes. 
Therefore, the basic equations of perfect fluid can not include any dissipation. 
Then we notice that \emph{the local equilibrium assumption is inconsistent with the irreversible nature of dissipative phenomena}. 
In other words, dissipations can not exist in local equilibrium systems. 
On the other hand, recall that the classic laws of dissipations are also the phenomenologies assuming the local equilibrium. 
Hence the classic laws lead inevitably an unphysical conclusion; an infinitely fast propagation of dissipations. 
While the infinitely fast propagation may be harmless to Newtonian theories, however, in relativistic theories, it gives rise to a serious problem; the violation of causality. 
The classic laws of dissipations can not be accepted as basic laws of relativistic dissipations. 
(See~\cite{ref:eit} for details of the violation of causality by the local equilibrium assumption.)

From the above, it is recognized that we should abandon the local equilibrium assumption in order to obtain a consistent dissipative hydrodynamics. 
Therefore the idea of \emph{local \underline{non}-equilibrium} is necessary, which means that the fluid element is in a non-equilibrium state. 
A physically consistent phenomenology of dissipative hydrodynamics, which is based on the local \emph{non}-equilibrium idea, is already formulated. 
It is called the \emph{Extended Irreversible Thermodynamics} (EIT). 
As precisely explained in~\cite{ref:eit}, because of the local non-equilibrium idea, EIT describes the entropy production inside each fluid element, which results in a finite speed of propagation of dissipations in both non-relativistic and relativistic situations. 
The relativistic EIT is a causally consistent dissipative hydrodynamics~\cite{ref:israel,ref:is,ref:hl_causality,ref:hl_applicable}.

Although the EIT is a dissipative ``hydrodynamics'', it is called ``thermodynamics''. 
This name puts emphasis on the replacement of local equilibrium idea with local \emph{non}-equilibrium one, which is a revolution in thermodynamic treatment of fluid element. 
The terminology ``EIT'' is found in~\cite{ref:eit} which is developed by experts in non-equilibrium physics. 
Contrary, the original works of relativistic dissipative hydrodynamics~\cite{ref:israel,ref:is,ref:hl_causality,ref:hl_applicable} put emphasis not on the thermodynamic revolution but on the preservation of causality. 
Here let us dare to use the term ``EIT'' since the local non-equilibrium nature of dissipative fluid is the physical origin of preservation of causality.

Before showing the basic equations of EIT, we list the basic quantities;
\eqabnon
 u^{\mu}(x) &:& \mbox{velocity field of fluid (four-velocity of fluid element)}\\
 \rho(x)\,\,\Bigl(\,=\frac{1}{V(x)}\,\Bigr)
            &:& \mbox{mass density ($\rho$) and specific volume (volume per unit mass, $V$)}\\
 \ene(x)    &:& \mbox{non-equilibrium specific internal energy (internal energy per unit mass)}\\
 \pne(x)    &:& \mbox{non-equilibrium pressure}\\
 \Tne(x)    &:& \mbox{non-equilibrium temperature}\\
 q^{\mu}(x) &:& \mbox{heat flux vector}\\
 \Pi(x)     &:& \mbox{bulk viscosity}\\
 \sv\,^{\mu\nu}(x) &:& \mbox{shear viscosity tensor}\\
 g_{\mu\nu}(x) &:& \mbox{spacetime metric}
\eqaenon
Here $x$ denotes the dependence on spacetime point. 
All of the above quantities except for $u^{\mu}$ and $g_{\mu\nu}$ are the \emph{non-equilibrium thermodynamic state variables} which characterize the non-equilibrium state of each fluid element. 
The non-equilibrium state variables are classified into two categories: 
The quantities $\ene$, $\pne$, $\Tne$ and $\rho$ are the state variables which exit even at the local equilibrium limit, while the quantities $q^{\mu}$, $\Pi$ and $\sv\,^{\mu\nu}$ are the state variables which should vanish at the local equilibrium limit. 
We call the first category ($\ene$, $\pne$, $\Tne$, $\rho$) the \emph{non-equilibrium scalars}, and call the second category ($q^{\mu}$, $\Pi$, $\sv\,^{\mu\nu}$) the \emph{dissipative fluxes}. 
The dissipative fluxes are the origin of dissipative phenomena and make the fluid being non-equilibrium. 
The suffix ``ne'' for non-equilibrium scalars denotes ``non-equilibrium'', and the variables with this suffix have different value from an equilibrium case. 
Note that the definition of $\rho$ (or $V$) is the same for both equilibrium and non-equilibrium cases; to count the number of composite particles or measure the mass per unit volume. 
So the suffix ``ne'' is not given to $\rho$ and $V$.

As the basic assumption of EIT, it is required that the dissipative fluxes are \emph{independent} non-equilibrium state variables. 
Furthermore, concerning the non-equilibrium scalars, it is also assumed that, as in the ordinary equilibrium thermodynamics, the number of independent non-equilibrium scalars is two for \emph{closed} systems which conserve the number of composite particles, and three for \emph{open} systems in which the number of composite particles changes. 
The remaining state variables are expressed as functions of the independent variables through the equations of state, e.g. the non-equilibrium specific entropy $\sne$ is a function of independent state variables $\sne = \sne(\ene,V,q^{\mu},\Pi,\sv\,^{\mu\nu})$, where $\ene$ and $V$ are chosen as independent non-equilibrium scalars.

Here note the fact that EIT, at least in its present status, is not necessarily applicable to any non-equilibrium state of dissipative fluid~\cite{ref:eit}. 
The consistent basic equations of EIT can be formulated for sufficiently weak dissipative fluxes, which means that the strength of non-equilibrium nature should not be so strong. 
For example, it is shown in~\cite{ref:hl_applicable} that the EIT preserves causality of heat flow if $\sqrt{q^\mu q_\mu}/\ene \lesssim 0.08$ for non-viscous fluid ($\Pi = 0$, $\sv\,^{\mu\nu} = 0$) under stationary and homogeneous (spatially one-dimensional) condition. 
We can recognize that the amount of energy transported by dissipative fluxes should be less than a few percent of the internal energy which includes mass energy. 
This efficiency of dissipative energy transfer is larger than the efficiency of hydrogen burning in a star ($\sim O(10^{-3})$ ). 
Therefore, although the EIT is applicable to a dissipative fluid whose local non-equilibrium states are not so far from local equilibrium states, it is expected that the EIT is applicable to many astrophysical systems.

In the present status of EIT, the above restriction is reflected in the equations of state. 
The equations of state, e.g. $\sne(\ene,V,q^{\mu},\Pi,\sv\,^{\mu\nu})$, are expanded up to second order of dissipative fluxes about the \emph{fiducial equilibrium state}. 
Here the \emph{fiducial equilibrium state} is defined as an equilibrium state of fluid element of imaginary perfect fluid possessing the same fluid velocity $u^{\mu}(x)$ and mass density $\rho(x)$ with our actual dissipative fluid. 
Then, as explained in~\cite{ref:eit}, the evolution equations of dissipative fluxes are obtained under two requirements; (i)~positivity of entropy production rate at each fluid element, and (ii)~consistency with experimentally determined phenomenology of relaxation processes of dissipations. 
The resultant evolution equations are;
\eqab
\label{eq:heat}
 \tau_{\rm h}\,\cd{q}\,^{\mu}
 &=&
  - q^{\mu} - \lambda T\,\cd{u}\,^{\mu} + \tau_{\rm h}\,(q^{\nu}\cd{u}_{\nu})\,u^{\mu}
  - \lambda\,\Delta^{\mu\nu}\,\Bigl(\, 
     T_{,\nu} - T^2\,\bigl(\,\inthb\,\Pi_{,\nu}
                           + \inths\,\sv\,^{\,\,\,\alpha}_{\nu\,\,\,\,;\alpha} \,\bigr)
    \Bigr) \\
 \tau_{\rm b}\,\cd{\Pi}
 &=&
  - \Pi - \zeta\,u^{\mu}_{\,\,\,;\mu} + \beta_{\rm hb}\,\zeta\,T\,q^{\mu}_{\,\,\,;\mu} \\
\label{eq:shear}
 \tau_{\rm s}\,\bigl(\,\sv\,^{\mu\nu}\bigr)^{\cdot}
 &=&
  - \sv\,^{\mu\nu} + 2\,\tau_{\rm s}\,\cd{u}_{\alpha}\,\sv\,^{\alpha (\nu}\,u^{\nu)}
  - 2\,\eta\,\Bigl[\, u^{\mu;\nu} - T\,\inths\,q^{\mu;\nu} \,\Bigr]^{\circ} \,,
\eqae
where $T$ is the temperature of fiducial equilibrium state. 
Here, definitions of mathematical symbols are; $\cd{Q} \defeq u^{\mu}Q_{;\mu}$, $\Delta^{\mu\nu} \defeq u^{\mu} u^{\nu} + g^{\mu\nu}$ and $[A_{\mu\nu}]^{\circ} \defeq \Delta^{\mu\alpha}\Delta^{\nu\beta}A_{(\alpha\beta)} - (1/3)\Delta^{\mu\nu}\Delta^{\alpha\beta}A_{\alpha\beta}$.
And the meanings of phenomenological coefficients are; $\tau$'s are relaxation times of dissipative fluxes, $\lambda$ is heat conductivity, $\zeta$ and $\eta$ are respectively bulk and shear viscous rates, and $\inthb$ and $\inths$ are interaction coefficients between $q^{\mu}$ and $\Pi$ or $\sv\,^{\mu\nu}$. 
For example, $\inths$ means that the heat flow arises in a shear viscous flow, also shear viscosity arises in a flow with heating. 
Also $\inthb$ means the same between $q^{\mu}$ and $\Pi$. 
The values of those phenomenological coefficients should be determined by some experiments. 
Furthermore note that, because dissipative phenomena are not time reversal, these evolution equations of dissipative fluxes are also not time reversal.

In reference~\cite{ref:is} which is one of the original works of relativistic EIT, the \emph{molecular viscosity} are assumed as the physical origin of viscosities $\Pi$ and $\sv\,^{\mu\nu}$. 
However in Newtonian accretion disk theories~\cite{ref:disk}, the angular momentum in a disk is transported by the \emph{turbulent viscosity}. 
It seems to be appropriate to keep various possibility of the physical origin of viscosities. 
Hence, we treat the viscosities $\Pi$ and $\sv\,^{\mu\nu}$ as \emph{phenomenological} variables, whose physical origin is not specified. 
This means that the bulk and shear viscous rates, $\zeta$ and $\eta$, are left as undetermined parameter in the equations~\eqref{eq:heat} $\sim$~\eqref{eq:shear}. 
As mentioned in previous paragraph, $\zeta$ and $\eta$ are the phenomenological parameter whose values should be determined empirically.

The others of EIT's basic equations are given by the conservation of mass current, $(\rho\,u^{\mu})_{;\mu}=0$, and that of stress-energy-momentum tensor, $T^{\mu\nu}_{\quad;\nu}=0$ ;
\eqab
 \cd{\rho} + \rho\,u^{\mu}_{\,\,\,;\mu}
  &=& 0 \\
 \rho\,\left( \cd{\varepsilon} + p\,\cd{V} \right)
  &=&
  - q^{\mu}_{\,\,\,; \mu} - q^{\mu}\,\cd{u}_{\mu}
  - \left(\, \Pi\,\Delta^{\mu\nu} + \sv\,^{\mu\nu} \,\right)\,u_{\mu\,;\,\nu} \\
\label{eq:eom}
 \left(\, \rho\,\varepsilon + p + \Pi \,\right)\,\cd{u}\,^{\mu}
  &=&
  - \cd{q}\,^{\mu} + q_{\alpha}\,\cd{u}\,^{\alpha}\,u^{\mu} - u^{\alpha}_{\,\,\,; \alpha}\,q^{\mu}
  - q^{\alpha}\,u^{\mu}_{\,\,\,; \alpha}
  - \Delta^{\mu\alpha}\,
    \left(\, (p+\Pi)_{, \alpha} + \sv\,^{\,\,\,\beta}_{\alpha\,\,\,; \beta} \,\right) \,,
\eqae
where $\varepsilon$, $p$ and $T$ are the state variables of fiducial equilibrium state. 
The above equations~\eqref{eq:heat}~$\sim$ \eqref{eq:eom} and the Einstein equation are the basic equations of EIT.

Finally let us make a comment of which the original works of EIT~\cite{ref:eit,ref:israel,ref:is,ref:hl_causality,ref:hl_applicable} are not aware: 
The EIT can not be applied to radiation fluids as shown in~\cite{ref:rad}. 
Non-equilibrium radiation fluids need a special treatment different from the other matters. 
However, although the inconsistency of EIT with radiation fluid has been revealed, no satisfactory non-equilibrium phenomenology of radiation fluid exits at present.

\section{Dissipative Accretion Flow onto a Schwarzschild Black Hole}

As a preliminary report, we show two theorems, without precise proof, on dissipative flows around a Schwarzschild black hole. 
Both theorems can be proven by solving the basic equations of EIT without solving the Einstein equation and fixing the metric on Schwarzschild spacetime. 
The use of fixed background metric means that our analysis does not include the self-gravity of the dissipative fluid. 
Therefore the theorems shown below are applicable to non-self-gravitating dissipative flows, i.e. to the dissipative fluid whose total energy is sufficiently less than the mass of central black hole, however ``non-self-gravitating'' dissipative fluid is usually assumed in Newtonian accretion disk theories~\cite{ref:disk}.

For the first, let us consider the spherically symmetric dissipative accretion flow onto a Schwarzschild black hole:
\begin{thm}
Bulk and shear viscosity do not vanish in stationary spherically symmetric dissipative accretion flows onto Schwarzschild black hole.
\end{thm}
Naively one may think that the shear viscosity should vanish in spherical accretion flow. 
However theorem~1 denies this naive sense. 
Here note that the Navier-Stokes law also allows the existence of shear viscosity in spherical flow~\cite{ref:ray}. 
The difference between EIT and Navier-Stokes is that, while a flow wit only bulk (or shear) viscosity is possible in Navier-Stokes law, but such flow in EIT has an unphysical divergence; the density or speed of accretion diverges in the region far from black hole. 
Hence, in EIT, the bulk and shear viscosities do not arise separately, and they interact always. 
This result about viscosity may be a significant property of EIT in contrast with Navier-Stokes law.
Now I am trying to extend this theorem to a non-stationary spherical dissipative flow on general non-stationary spherical spacetime which includes the self-gravity of dissipative fluid.

For the second, let us consider the stationary toroidal dissipative flow around a Schwarzschild black hole:
\begin{thm}
For stationary toroidal flow with no mass accretion onto black hole, any flow of finite temperature is impossible, but only a rigid toroidal rotation of zero-temperature is possible.  
The fluid velocity and dissipative fluxes of the possible zero-temperature flow are
\eqb
\label{eq:u}
 u^t = \frac{1}{\sqrt{f(r)-\Omega^2\,r^2\,\sin^2\theta}} \quad,\quad
 u^r = u^\theta = 0 \quad,\quad
 u^\phi = \Omega\,u^t \qquad\mbox{and}\quad q^{\mu}=\Pi=\sv\,^{\mu\nu}=0 \,,
\eqe
where $(t,r,\theta,\phi)$ is the Schwarzschild coordinate, $f(r) \defeq 1 - 2M/r$ with $M$ as the black hole mass, and $\Omega$ is the angular velocity measured by a rest observer. 
Since $\Omega$ is constant and dissipations vanish even when they are switched on, the velocity $u^{\mu}$ in equation~\eqref{eq:u} expresses a rigid rotation. 
And the thermodynamic state variables of this flow are
\eqb
\label{eq:state}
 T = p = \varepsilon = 0 \quad,\quad
 \rho(r,\theta) = \mbox{arbitrary in the region $f(r)>\Omega^2\,r^2\,\sin^2\theta$}
\eqe
Here recall that the internal energy $\varepsilon(x)$, for a fluid element at $x$, includes the mass energy, kinetic energy and the potential energy by external gravity due to black hole. 
Therefore the vanishing internal energy $\varepsilon=0$ means that, for each fluid element, the potential energy due to black hole cancels out the kinetic and mass energies of fluid element.
\end{thm}
This theorem together with the third law of thermodynamics implies that any flow of finite temperature is non-stationary and/or non-toroidal, and never relaxes to the flow given in equations~\eqref{eq:u} and~\eqref{eq:state}. 
Hence, although a differentially rotating and stationary toroidal flow of finite temperature is possible for ``perfect'' fluid~\cite{ref:fm}, such stationary flow becomes unstable once the dissipations are switched on. 
Theorem~2 implies the dissipative instability of an exact solution of relativistic perfect fluid.

The dissipative instability, the EIT's effect, may describe a formation of out-flow from an accretion disk without introducing a magnetic field (so-called MRI), while the ``acceleration'' of out-flow to form a jet will be explained with magnetic fields. 
In (near) future, it is expected to apply the EIT to a plasma composing an accretion disk.


\end{document}